\documentclass[aps,superscriptaddress,floatfix,a4paper]{revtex4}
\usepackage{graphicx}
\usepackage{amsmath}
\usepackage{amssymb}
\usepackage{epsf,latexsym}
\usepackage{times}
\usepackage{epsfig}
\usepackage{citesort}

 \normalsize

\def\be{\begin{equation}}
\def\ee{\end{equation}}
\def\bea{\begin{eqnarray}}
\def\eea{\end{eqnarray}}
\def\l{\label}

\def\nn{{\nonumber}}
\def\lsim{\raise0.3ex\hbox{$\;<$\kern-0.75em\raise-1.1ex\hbox{$\sim\;$}}}
\def\gsim{\raise0.3ex\hbox{$\;>$\kern-0.75em\raise-1.1ex\hbox{$\sim\;$}}}

\def\ie{{\it i.e.}}
\begin{document}
\begin{flushright}
HIP-2009-27/TH
\end{flushright}

\title{New Physics contribution to $B \to K \pi$ decays in SCET}

\author{K. Huitu}
\affiliation{Department of Physics, and Helsinki Institute of Physics, P.O.Box 64, FIN-00014 University of
Helsinki, Finland}
\author{S. Khalil}
\affiliation{Centre for Theoretical Physics, The British
University in Egypt, El Sherouk City, 11837, Egypt\\
and Department of Mathematics, Ain Shams University,
Faculty of Science, Cairo, 11566, Egypt.}

\date{\today }

\begin{abstract}
We analyze the $5\sigma$ difference between the CP asymmetries of
the $B^0 \to K^+ \pi^-$ and $B^+ \to K^+ \pi^0$ decays within the
Soft Collinear Effective Theory.
We find that in the Standard Model, such a big difference cannot be
achieved.
We classify then the requirements for the possible New Physics models,
which can be responsible for the experimental results.
As an example of a New Physics model we study minimal supersymmetric
models, and find that the measured asymmetry can be obtained with
non-minimal flavor violation.
\end{abstract}
\maketitle

\section{ \bf Introduction}

The first observation of CP violation was in the neutral kaon
system in 1964, which was consistent with
Cabibbo--Kobayashi–-Maskawa (CKM) mechanism and with its
simplicity. In the last years, experiments at $B$-factories have
established CP violation in $B^0_d$ decay. Although the Standard
Model (SM) is able, till now, to account for the CP violating
experimental results, CP violation is one of the most interesting
aspects and unsolved mysteries of the SM. There are strong hints
of additional sources of CP violation beyond the phase in the CKM
mixing matrix. The strongest motivation for this suggestion is
that the strength of CP violation in SM is not sufficient to
explain the cosmological baryon asymmetry of our universe.
Therefore, it is expected that a sizeable contribution from New
Physics (NP) to CP violation in $B$-meson decays may be probed.

Indeed, there are some discrepancies between the SM expectations
and the experimental measurements of the following parameters:
${\rm sin}2\beta_s$ extracted from the mixing CP asymmetry in $B
\to J/\psi \phi$ decay~\cite{Bona:2008jn}, sin$2\beta$ extracted
from the mixing CP asymmetry in $B \to K \phi$ and $B \to K \eta'$
decays~\cite{HFAG}, and the direct CP asymmetries of $B\to K \pi$
decays. Of these, the $B \to K\pi$ anomaly remains a potential
hint for NP that emerges from rare $B$ decays. The current world
averages for the branching ratios (BRs) and CP asymmetries of
$B\to K \pi$ \cite{HFAG} are summarized in Table 1.
\begin{table}[t]
\begin{center}
\begin{tabular}{|c|c|c|}
  \hline
  \hline
  \hspace{1cm} Decay channel \hspace{1cm} &  \hspace{1cm} $BR \times 10^{-6}$ \hspace{1cm} & \hspace{1cm}
$A_{\scriptscriptstyle CP}$ \hspace{1cm} \\
  \hline
  \hline
  $K^+ \pi^-$ & $19.4\pm 0.6 $ & $ -0.098\pm 0.012$ \\
  $K^+ \pi^0$ & $12.9 \pm 0.6$ & $0.050\pm 0.025$ \\
  $K^0 \pi^+$ & $23.1 \pm 1.0 $ & $0.009\pm 0.025$ \\
  $K^0 \pi^0$ & $9.8 \pm 0.6 $ & $-0.01\pm 0.1$ \\
  \hline
\end{tabular}
\vskip 0.25cm \caption{The latest average results for the BRs and CP
asymmetries of $B \to K \pi$ decays.}\label{asymtable}
\end{center}
\end{table}
These results confirm the existence of a non-vanishing difference
between the asymmetries of $B^+ \to K^+ \pi^0$ and $B^0 \to K^+
\pi^-$ beyond $5 \sigma$:
\begin{eqnarray}
{\cal A}_{CP}(B^+ \to K^+ \pi^0) - {\cal A}_{CP}(B^0 \to K^+ \pi^-
)
=(14.8\pm 2.7)\%~. ~~~~ \label{difference}
\end{eqnarray}

It is well known that within the SM, all CP violating processes
should be accommodated by the single phase of the CKM, which is
the only source of CP violation in the quark sector. This implies
tight relations among the CP asymmetries of different processes,
which allow stringent tests of the SM, and may therefore lead to
the discovery of NP. Indeed, the SM results for the CP asymmetries
of $B \to K \pi$, with naive factorization or "improved" BBNS QCD
factorization \cite{beneke} (QCDF), indicate that the above
mentioned two asymmetries are essentially equal
\cite{khalil:2005qg}. This inconsistency is known as $B \to K \pi$
puzzles and has been considered as a possible hint for physics
beyond the SM, with new source of CP violation. There has been
tremendous work over the last few years in order to understand
this puzzle of CP asymmetries in $B \to K\pi$ decays.

In this paper, we perform a detailed analysis for the CP
asymmetries and branching ratios of $B \to K \pi$ decays in the
framework of SCET \cite{Bauer:2000ew,Bauer:2005kd}. In
Ref.\cite{Bauer:2005kd}, the SM contributions to the branching
ratios and the CP asymmetries of $B \to K \pi$ have been studied
in the frame of the SCET. It was concluded that a small CP
asymmetry for $B^+ \to K^+ \pi^0$ is predicted and the large
discrepancy between the CP asymmetries of $B^0 \to K^+ \pi^-$ and
$B^+ \to K^+ \pi^0$ is difficult to explain in the SM with SCET
and a possible new source of new physics in order to account for
these results. Motivated by this conclusion and also by the fact
that the difference between these two asymmetries is now reached
$5 \sigma$, we study the new physics, in particular supersymmetry,
contributions to these processes and analyze the conditions that
may allow producing the recent experimental results.

The SCET provides a systematic and elegant method for calculating
$B$ decays with several relevant energy scales
\cite{Bauer:2000ew,Bauer:2005kd,Bauer:2000yr,Chay:2003zp,Chay:2003ju,Jain:2007dy,Bauer:2002aj}.
It is based on the fact that the decay of heavy hadrons to highly
energetic light hadrons includes three distinct energy scales: the
hard energy scale $\sim m_b $, the hard collinear scale $\sim
\sqrt{m_b \Lambda_{QCD}}$ and the hadronic soft scale $\sim
\Lambda_{QCD}$. Thus, the matching of the weak effective
Hamiltonian into the corresponding SCET gauge invariant operators
requires two step matching \cite{Bauer:2002aj}.  First the
effective weak Hamiltonian is matched to the corresponding weak
Hamiltonian in what is called SCET$_I$, by integrating out at the
hard modes with momentum  of order $m_b$. Second, the SCET$_I$
weak Hamiltonian is matched onto the weak Hamiltonian SCET$_{II}$
by integrating out the hard collinear modes with $p^2\sim
m_b\Lambda_{QCD}$. Accordingly, the SCET is improving the
factorization, obtained from expansion in powers of
$\Lambda_{QCD}/m_b$, by generalizing it to allow each of the above
mentioned scales to be considered independently. We will show
explicitly that, as in the QCDF approach, the SM results for CP
asymmetries of $B \to K \pi$ in SCET are typically not consistent
with the observed measurements. This confirms the conclusion that
NP is required in order to accommodate the experimental
measurements of $B \to K \pi$ CP asymmetries. We will analyze the
type of NP needed to resolve $B \to K \pi$ puzzle and show that it
must induce new source of CP violation. As an interesting example
of NP, we consider the supersymmetric (SUSY) extension of the SM,
using the mass insertion approximation (MIA) in order to perform a
model independent analysis.

It is important to note that in order to have significant CP
violating effects from SUSY contributions without exceeding the
experimental limits of the Electric Dipole Moment (EDM) of
electron and neutron, one should consider a SUSY model with
non-minimal flavor. In this class of models, like for instance
scenarios of non-universal trilinear couplings, there are new
sources of CP and flavor violation that may lead to significant
impacts on the CP asymmetries of $B \to K\pi$, without violating
the experimental limits of the electric dipole moment (EDM) of
electron or neutron \cite{Abel:2001vy}. It has been emphasized in
Ref. \cite{khalil:2005qg,Khalil:2009zf} that these phases are
crucial in providing a natural explanation for the $B \to K \pi$
puzzle. Indeed, this new source of SUSY CP violating phases
induces CP violating phases associated with the electroweak
penguins, which are essential with large strong phase in order to
resolve the apparent discrepancies between the CP asymmetry of
$B^+\to K^+ \pi^0$ and $B^0\to K^+ \pi^-$.

The paper is organized as follows. In Section 2 we discuss $B\to K
\pi$ process in the SCET and present generic expressions for the
amplitudes in terms of the Wilson coefficients. Section 3 is
devoted for analyzing the SM contribution to the branching ratios
and CP asymmetries of $B\to K \pi$ decays. We show that the
branching ratios can be consistent with the experimental data if a
large charm penguin contribution is assumed. Nevertheless, the CP
asymmetries measurements cannot be accommodated. In Section 4 we
explore the NP effects and possible types of NP that may resolve
the puzzle of $B \to K \pi$. We emphasize that a generic feature
of any of this NP is that it must introduce a new source of CP
violation. In Section 5 we focus our discussion on SUSY extension
of the SM. We show that the gluino contribution to the electroweak
penguin plays a crucial role in resolving the $B \to K \pi$
puzzle. Finally we summarize our conclusions in Section 6.


\section{ \bf $ B\to {K \pi}$ in SCET }

The full effective weak Hamiltonian $H^{\Delta B=1}_{\rm eff}$ for
$\Delta S=1 $ transitions can be expressed via the operator
product expansion as
 \begin{eqnarray}
H^{\Delta B=1}_{\rm eff}&=& \frac{G_F}{\sqrt{2}} \sum_{p=u,c}
\lambda_p^{(s)} ~\left( C_1 Q_1^p + C_2 Q_2^p + \sum_{i=3}^{10}
C_i Q_i + C_{7\gamma} Q_{7\gamma} + C_{8g} Q_{8g} \right)+\left(
Q_i\rightarrow \tilde{Q_i},C_i\rightarrow\tilde{{C_i}}\right),~~
 \label{Heff}
\end{eqnarray}
where  $\lambda_p^{(s)}= V_{pb} V^{\star}_{p s}$, with $V_{ij}$
the unitary CKM matrix elements. $C_i \equiv C_i(\mu_b)$ are the
Wilson coefficients at low energy scale $\mu_b \simeq {\cal
O}(m_b)$. The operators $Q_i$ can be found in
Ref.~\cite{Buchalla:1995vs}. The operators $Q^p_{1,2}$ refer to the
current-current operators, $Q_{3-6}$ to the QCD penguin operators,
and $Q_{7-10}$ to the electroweak penguin operators, while
$Q_{7\gamma}$ and $Q_{8g}$ are the electromagnetic and the
chromomagnetic dipole operators, respectively.  The operators
$\tilde{Q_i}$ are obtained from $Q_i$ by the chirality exchange.
It is important to note that the electroweak penguins and the
electromagnetic penguin are the only source of isospin violation,
which is indicated by the $K \pi$ puzzle.

The calculation of $B\to K \pi$ decays involves the evaluation of
the hadronic matrix elements of related operators in the effective
Hamiltonian, which is the most uncertain part of this calculation.
In the limit in which $m_b \gg \Lambda_{QCD}$ and neglecting QCD
corrections in $\alpha_s$, {\it i.e.} in the Naive Factorization (NF) approach, the
hadronic matrix elements of $B$ decays into $K$ and $\pi$ can be
factorized as%
\be%
\langle K \pi \vert Q_i \vert B \rangle_{NF} = \langle K \vert j_1
\vert B \rangle \times \langle \pi \vert j_2 \vert 0 \rangle +
\langle \pi \vert j_1
\vert B \rangle \times \langle K \vert j_2 \vert 0 \rangle, %
\ee %
where $j_{1,2}$ represent bilinear quark currents of local
operator $Q_i$. Therefore, the hadronic matrix element can be
usually parameterized by the product of the decay constants and
the transition form factors.

In QCDF the hadronic matrix element
for $B \to K \pi$ in the heavy quark limit $m_b \gg \Lambda_{QCD}$
can be written as %
\be %
\langle K \pi \vert Q_i \vert B \rangle_{QCDF} = \langle K \pi
\vert Q_i \vert B \rangle_{NF} ~ \left[ 1 + \sum_n r_n \alpha_s^n
+ {\cal O}\left(\frac{\Lambda_{QCD}}{m_b}\right)\right].%
\ee%
It is clear that in QCDF, the higher order corrections in
$\alpha_s$ break the simple factorization. These corrections can
be calculated systematically in terms of short-distance
coefficients and meson light-cone distribution functions. However,
it turns out that the calculation of the hard spectator
interactions and the annihilation amplitude suffer from end-point
divergences in this factorization approach. The divergences are
parameterized by complex parameters with magnitudes less than one
and unconstrained phases. Such parameters are the main source of
large theoretical uncertainties in the QCDF mechanism.

The SCET is an interesting framework to study the factorization at
hard ${\cal O}(m_b)$ and hard-collinear ${\cal O}(\sqrt{m_b
\Lambda_{QCD}})$ scales. The SCET Lagrangian is obtained at tree
level by expanding the full theory Lagrangian in powers of $
\lambda = \Lambda_{QCD}/m_b$. This would allow to prove or
disprove the factorization to all orders in the strong coupling
constant for some B decays into light and energetic particles.
Many theoretical works have been done in the SCET, in particular
the matching of ${\rm QCD} \to {\rm SCET}_I \to {\rm SCET}_{II}$
and the derivation of the amplitudes for the $B$ decay into light
mesons
\cite{Bauer:2000ew,Bauer:2005kd,Bauer:2000yr,Chay:2003zp,Chay:2003ju,Jain:2007dy}.
For $B \to K \pi$, the SCET amplitude can be written as %
\bea%
{\cal A}_{B\to K\pi}^{SCET} &=&
-i\langle K\pi|H_{eff}^{SCET}| B \rangle \nn\\
 &=& {\cal A}_ {B\to K\pi}^{LO} +  {\cal A}_{B\to K\pi}^ {\chi} +
 {\cal A}_{B\to K\pi}^{ann} +  {\cal A}_{B\to K\pi}^{c.c} %
\eea %
where ${\cal A}_{B\to K\pi}^{LO}$ denotes the leading order
amplitude in the expansion $1/m_b$ (including correction of order
$\alpha_s$), $ {\cal A}_{B\to K\pi}^ {\chi}$ denotes the chirally
enhanced penguin amplitude, ${\cal A}_{B\to K\pi}^{ann}$ denotes
the annihilation amplitude and $ {\cal A}_{B\to K\pi}^{c.c}$
denotes the long distance charm penguin contributions.

The  leading order amplitude, ${\cal A}_ {B\to K\pi}^{LO}$, is
given by
\begin{eqnarray}
{\cal A}^{LO}_{B\rightarrow K\pi}&=&\frac{G_F
m_B^2}{\sqrt{2}}\left[f_{K}\left(\int^1_0 dudzT_{KJ}(u,z)\zeta^{B
\pi}_J(z)\phi_{K}(u) +\zeta^{B \pi}\int^1_0
 du T_{K \zeta}(u)\phi_{K}(u)\right)+ \left(K \leftrightarrow
 \pi\right)\right].~~ \label{amp1}
 \end{eqnarray}
The hard kernels $T_{(K,\pi)\zeta}$ and $T_{(K,\pi) J}$ are
calculable in terms of the Wilson coefficients $C_i$ and can be
found in Ref.~\cite{Bauer:2004tj}. The parameters $\zeta^{B
(K,\pi)}$, $\zeta_J^{B (K,\pi)}$ are treated as hadronic
parameters that can be determined through the fit to the non
leptonic decay data. The current data can be used to determine
$\zeta^{B \pi}$, $\zeta_J^{B \pi}$. However these data are not
sufficient to determine $\zeta^{B K}$ and $\zeta_J^{B K}$and hence
we assume $\zeta_J^{B K}=\zeta_J^{B \pi}$ and $\zeta^{B
K}=\zeta^{B \pi}$ in the limit of exact $SU(3)$. One may expect
about $10\%-20\%$ deviation in the values of these parameters
in case of $SU(3)$ breaking.

It is important to note that as long as the logarithms of the
ratios of the hard scale ($m_b$) to the soft-collinear ($\Lambda
m_b$) and soft ($\Lambda$) scales are not re-summed, the QCDF and
SCET factorization formulae are identical. Therefore,
Eq.(\ref{amp1}) for the expression of $A_{B\to K\pi}^{LO}$
includes as well the end-point singular contribution mentioned
above in QCDF scheme. In fact, the form factors $\xi^{B\pi}_J(z)$,
which are extracted from the data, could be expressed as end-point
singular convolutions between the pion and $B$-meson light-cone
wave functions.

Chiraly enhanced penguins amplitude $ {\cal A}_{B\to K\pi}^
{\chi}$ is generated through including  corrections of order
$\alpha_s(\mu_h)(\mu_M\Lambda/m_b^2)$ where $\mu_M$ is the chiral
scale parameter. $\mu_M$ is defined as the ratio of the squared
meson mass to the sum of its constituent quark masses. For kaons
and pions $\mu_M\sim{\cal O}$(2) GeV  and hence chiraly enhanced
terms can compete with the order $\alpha_s(\mu_h)(\Lambda/m_b)$
terms. The chiraly enhanced amplitude for  $B\to K \pi$ decays is
given by
\begin{eqnarray}%
\label{chienhanced} %
A^\chi_{\bar B\to K \pi} &=& \frac{G_F m_B^2}{\sqrt2}  \bigg\{-
\frac{\mu_{K}  f_{K}}{3m_B} \zeta^{B\pi} \int_0^1 du~
R_{K}(u)\phi_{pp}^{K}(u) -  \frac{\mu_{K} f_{K}}{3m_B} \int_0^1 du
dz~ R_{K}^{J}(u,z) \zeta_J^{B\pi}(z) \phi_{pp}^{K}(u) \nonumber\\
& -& \frac{\mu_{\pi}f_{K}}{6 m_B}\int_0^1 du dz~ R_{K}^{\chi}(u,z)
\zeta_{\chi}^{B\pi}(z) \phi^{K}(u) +(K\leftrightarrow \pi)
\bigg\}.
\end{eqnarray}
The hard kernels $R_{K},R_{\pi},R_{K}^J,R_{\pi}^J,R_{K}^{\chi}$
and $ R_{\pi}^{\chi}$ depend also on $C_i$, as shown in
\cite{Jain:2007dy}.

Annihilation amplitudes ${\cal A}_{B\to K\pi}^{ann}$ have been
studied in
Ref.~\cite{Keum:2000ph,Lu:2000em,Beneke:2001ev,Kagan:2004uw}. In
the framework of SCET, the annihilation contribution becomes
factorizable and  real at leading order, ${\cal
O}(\alpha_s(m_b)\Lambda/m_b)$. Complex annihilation contributions
may occur at higher order, ${\cal O}(\alpha_s^2(\sqrt{{m_b
\Lambda}})\Lambda/m_b)$\cite{Arnesen:2006vb}.  In our numerical
analysis, we will not include the contributions from penguin
annihilations, since they are real, at the order we consider, and
are quite small with large uncertainty
\cite{Arnesen:2006vb,Jain:2007dy}. It is worth mentioning that
there are some question marks related to the SCET result for
$A^{ann}_{B\to K \pi}$. It is expected that the approach adopted
in computing the LO expression may lead to a divergent
annihilation contribution, which therefore requires a
re-introduction of complex parameter as in QCDF. This discussion
is beyond the scope of this paper, specially in case of neglecting
the annihilation amplitude.

The long distance  charm penguin amplitude $ {\cal A}_{B\to
K\pi}^{c.c}$ is given as follows \be {\cal A}_{B\to
K\pi}^{c.c}=|{\cal A}_{B\to K\pi}^{c.c}|e^{i \delta_{cc}} \ee
where $\delta_{cc}$ is the strong phase of the charm penguin. The
modulus and the phase of the charm are fixed, through the fitting
with non leptonic decays, namely $B\rightarrow \pi\pi$, assuming
${\cal A}_{B\to K\pi}^{c.c}={\cal A}_{B\to \pi\pi}^{c.c}$, as
follows \cite{Williamson:2006hb}:%
\begin{eqnarray}
&&\vert A_{c.c} \vert = (46 \pm 0.8) \times 10^{-4}, ~~~~ ~~~
\delta_{c.c} = 156^o \pm 6^o.%
\label{Acc}
\end{eqnarray}
 The charm penguin can be considered as one of the main
differences between SCET and QCDF. In QCDF, it is factorized in
the limit of $1/m_b$. However, in SCET, since $m_c \sim m_b/2$
there may be configurations, where charm penguin implies long
distance effect. Thus, it has been parameterized and fitted from
the data. It is also worth noting that in SCET the charm penguin
is the main source of strong phases in the decay amplitudes. All
strong phases for other terms vanish at the leading order.

The unitarity of the CKM matrix allows to write the amplitude of any $B$-decay as
$A = \lambda_u^{(f)} A_u + \lambda_c^{(f)} A_c$, where $\lambda_p^{(f)}=V_{pb}^* V_{pf}$.
Thus, one can generally parameterize the contributions to the
amplitudes of $B \to K \pi$ as follows:
\begin{eqnarray}
A(B^+\to K^0 \pi^+)&=&\lambda_u A  + \lambda_c P, \nonumber\\
\sqrt{2} A(B^+\to K^+ \pi^0 ) &=& \lambda_u \left( T + C + A
\right) + \lambda_c \left(P + P_{EW} \right),\nonumber\\
A(B^0\to K^+ \pi^-)&=&\lambda_u T + \lambda_c \left( P + P_{EW}^C \right), \nonumber\\
\sqrt{2} A(B^0\to K^0 \pi^0 ) &=& \lambda_u C - \lambda_c \left(P
- P_{EW} + P_{EW}^C\right). \label{parametrization1}
\end{eqnarray}
where the real parameters: $T,C,A, P, P_{EW}$, and $P_{EW}^C$
represent a colored allowed tree, a color suppressed tree,
annihilation, QCD penguin, electroweak penguin, and suppressed
electroweak penguin diagrams, respectively. The four $B\to K \pi$
decay amplitudes are related by the following isospin relation:
\begin{equation}
\sqrt{2} A(B^0\to K^0 \pi^0) + A(B^+\to K^0 \pi^+) - \sqrt{2} A(B^+\to K^+ \pi^0)
+ A(B^0\to K^+ \pi^-) =0.
\end{equation}
The explicit dependence of these parameters on the corresponding
Wilson coefficients can be found in
Ref.~\cite{Bauer:2000ew,Bauer:2000yr,Chay:2003zp,Chay:2003ju,Jain:2007dy}.
Fixing the experimental inputs and the SM parameters to their
center values, one finds the following dependence of these
parameters  on the Wilson coefficients (at NLO in $\alpha_s$
expansion of SCET + $SU(3)$ flavor symmetry):
\begin{eqnarray}
\hat{A} &=& -(0.0003 + 0.0005 i) C_1 - 0.0134 C_{10}
 + (0.0233 - 0.0009 i) C_3 + 0.0268 C_4 + 0.0113 C_5 + 0.034 C_6\nn\\
&-& 0.0057 C_7 - 0.017 C_8  - (0.012 - 0.0005 i) C_9 - 0.0009 C_{8g},\nn\\
\hat{P}&=& (-0.0004 - 0.0003 i) C_1 - 0.013  C_{10} + (0.0234 -  0.0009 i) C_3 + 0.027 C_4 + 0.0113 C_5 + 0.034 C_6 \nn\\
&-& 0.006 C_7 - 0.017 C_8 -(0.012 -0.0005i) C_9 - 0.0009 C_{8g} -(0.004 -0.002 i),\nn\\
\hat{P}_{EW}^C&=&  0.017 C_7 + 0.051 C_8 +(0.035 - 0.0014i) C_9 +0.04 C_{10}, \nn\\
\hat{P}_{EW}&=& -0.016 C_7+(0.056 - 0.0014 i) C_8 +(0.068 - 0.0014 i)C_9 +(0.066 - 0.0013 i) C_{10},\nn\\
\hat{C}&=&  (0.017 - 0.0004 i) C_1 + (0.039 - 0.001 i) C_{10}
+  0.022 C_2- (0.023 - 0.0009 i) C_3 - 0.027 C_4 - 0.011 C_5\nn\\
&&- 0.034 C_6 - 0.027 C_7 + (0.022- 0.0014 i) C_8 +0.0009 C_{8g}+ (0.045 - 0.0005 i) C_9 , \nn\\
\hat{T}&=& (0.027 - 0.001 i) C_1 +  0.027 C_{10} + (0.023 - 0.001 i) C_2
 + (0.024 -   0.001 i) C_3 +0.027 C_4 + 0.011 C_5 \nn\\
&&+ 0.034 C_6 + 0.0113 C_7  + 0.034 C_8 -  0.0009 C_{8g} + (0.024
- 0.001 i) C_9,
\end{eqnarray}
where $\hat{X}$ is defined as $\hat{X}=\sqrt{2}X/ G_F m_B^2$ with
$X\equiv A, T,C,P,P_{EW}, P_{EW}^C$. The above results correspond
to the total amplitudes including the chirally enhanced penguin
with inclusion of the charm penguin as a nonperturbative contribution
fitted from the experimental data. Note that the charm penguin
contributes only to the QCD penguin $P$, and it is fixed from the
the data of $B\to \pi \pi$ processes.

Here few comments are in order: $(i)$ At leading order, the only
source of the strong phases is the charm penguin, however at next
leading order correction, small strong phases may emerge.
$(ii)$ In the combined SCET+$SU(3)$, one finds that $C \sim T$,
hence there is no color suppression. $(iii)$ There is no
undetermined strong phase in the amplitudes $T,C, P_{EW},
P_{EW}^C$, unlike the QCDF. Thus, the relative sign of CP
asymmetries is predicted. $(iv)$ The amplitudes $P_{EW}$ and
$P_{EW}^C$ receive contributions through the electroweak penguin
operators $O_{7-10}$. Unlike the gluonic penguins, the electroweak
($\gamma$- and $Z$- mediated) penguins distinguish the up from the
down quark pairs in the final state. Therefore, if they are not
suppressed, they may account for the difference between the CP
asymmetries in the two isospin related decays of Eq.~(\ref{difference}).

\section{ \bf SM contribution to the CP asymmetry of {$ B\to K \pi$ }}
In this section we reappraise the SM predictions for the CP
asymmetries of $B \to K \pi$ decays in SCET \cite{Bauer:2005kd}.
In the NDR scheme taking $\alpha_s(m_Z)=0.118$, $m_t=170.9\,{\rm
GeV}$, and
$m_b=4.7\,{\rm GeV}$, the Wilson coefficients are given by:%
\bea %
C_{1-10}(m_b) &=& \{ 1.078\,,  -.177\,, .012\,, -.0335\,, .0095\,,
-.040 \,,  1 \!\times\! 10^{-4} \,,  4.2 \!\times\! 10^{-4} \,,
-9.7 \!\times\! 10^{-3} \,,
  1.9 \!\times\! 10^{-3} \},\nn \\
C_{7\gamma}(m_b)&=&-.316 \,, C_{8g}(m_b) =-0.149.
\eea %
As can be seen from these values, the SM contributions to the
electroweak penguins $C_7 - C_{10}$ are quite suppressed. Thus,
one expects that the EW penguins in the SM are negligible, hence
the $B\to K \pi$ asymmetries are  dominated by the QCD penguins,
which give universal contributions to the four decay channel.
Accordingly, it is expected that the SM results for the CP
asymmetries of different $B \to K \pi$ channels are very close.
Since the SM Wilson coefficients are real, one can rewrite the
amplitude of $B\to K \pi$ in Eq.~(\ref{parametrization1}) as:
\begin{eqnarray}%
\label{Kpigeneral}%
A(B^+ \to K^0  \pi^+ )& =& \lambda_c^{(s)} P \Big[ 1 +
 r_A e^{i(\delta_A- \gamma)} \Big]\,, \nn \\
 A(B^0 \to K^+ \pi^-) & =&  \lambda_c^{(s)} P \Big[ 1 +
\big( r^C_{EW} e^{i\delta^C_{EW}} + r_T e^{i(\delta_T-\gamma)}\big) \Big], \nn \\
 \sqrt{2} A(B^+ \to K^+ \pi^0)&=&  \lambda_c^{(s)} P \Big[ 1 +
 \big( r_{EW} e^{i\delta^{EW}} + r_T e^{i(\delta_T-\gamma)}
+ r_C e^{i(\delta_C-\gamma)}+ r_A e^{i(\delta_A-\gamma)}  \big)  \Big], \nn \\
\sqrt{2} A(B^0 \to K^0 \pi^0) &=& \lambda_c^{(s)} P \Big[ - 1 +
\big( r_{EW} e^{i\delta_{EW}} - r^C_{EW} e^{i\delta_{EW}^C} + r_C
e^{i(\delta_C-\gamma)} \big)
  \Big],
\label{SMamplitude1}
\end{eqnarray}%
where the parameters $\delta_{J}$, with $J$ stands for $T,C,A,
EW,EW^C$, are the CP conserving (strong) phase and $r_J$ are
defined as %
\be
  r_T e^{i\delta_T} =
     \bigg| \frac{\lambda_u^{(s)}}{\lambda_c^{(s)}} \bigg| \
     \frac{T}{P}  \,, ~~~  r_C e^{i\delta_C}
   = \Big| \frac{\lambda_u^{(s)}}{\lambda_c^{(s)}} \Big| \
     \frac{C}{P} \,,   ~~~  r_A e^{i\delta_A}
   = \Big| \frac{\lambda_u^{(s)}}{\lambda_c^{(s)}} \Big| \
     \frac{A}{P}\,,~~  r_{EW}\: e^{i\delta_{EW}} =
    \frac{P_{EW}}{P} , ~~~  r^C_{EW} \: e^{i\delta^C_{EW}} =
    \frac{P_{EW}^C}{P}.
    \label{r-SM}
 \ee  %
As can be seen from Eq.~(\ref{parametrization1}), $P$ is dominated
by the large charm penguin. Therefore, one finds that all the
above ratios are quite suppressed and also have one single strong
phase, which is essentially $\delta_{c.c}$. Namely, one obtains
the following results
\bea%
r_T e^{i\delta_T}&=&0.06~ e^{-2.91i},~~~ r_C e^{i\delta_C}=0.05~
e^{-2.92i},~~~ r_A e^{i\delta_A}=0.006~ e^{0.54i},\nonumber\\
r_{EW} e^{i\delta_{EW}}&=& 0.08~ e^{0.23i},~~~~ r^C_{EW}
e^{i\delta^C_{EW}}= 0.04~ e^{0.2i}.%
\eea
>From these results, one notices that in SCET the ratio between the
color-suppressed tree and color-allowed tree is enhanced, so
$\vert C/T \vert \sim 1$, unlike the corresponding ratio in QCDF.
This enhancement is due to the suppression of $T$, not because
enhancement of $C$. In this approach, one finds $r_T \sim r_C$ and
$r_{EW} \sim r_{EW}^C$, which means there is no color suppression.
However, even if color suppressed tree and electroweak penguin $(C
, P_{EW}^C)$ are enhanced and become of the order of the color
allowed tree and electroweak penguin $(T , P_{EW})$, it is not
possible to resolve the puzzle $B \to K \pi$ CP asymmetry in the
framework of the SM, due to a lack of CP violation as emphasized
in Ref.~\cite{khalil:2005qg}. Due to the dominance of $A_{c.c.}$
in $P$, hence $r_{J} \ll 1$, the following relation between the
amplitudes
of different channels is established:%
\be%
A_{K^0 \pi^+} \simeq A_{K^+ \pi^-} \simeq \sqrt{2} A_{K^+ \pi^0}
\simeq \sqrt{2} A_{K^0 \pi^0}. %
\ee%
The branching ratio of $B \to K \pi$  is given by%
\be%
BR(B\to K \pi) = \frac{1}{\Gamma_{tot}}\frac{\left[\left(M_B^2 -
(m_K + m_{\pi})^2\right)\left(M_B^2 - (m_K -
m_{\pi})^2\right)\right]^{1/2}}{16 \pi M_B^3}\Big[~
\vert A_{K \pi}\vert^2 + \vert \overline{A_{K\pi}}\vert^2 \Big]. %
\ee%
Therefore, the BRs also satisfy the relation: %
\be %
BR_{K^0 \pi^+} \simeq BR_{K^+ \pi^-} \simeq 2 BR_{K^+ \pi^0}
\simeq 2 BR_{K^0 \pi^0}, %
\ee%
which is consistent with the data given in Table 1. However, the
magnitude of the BR is sensitive to the value of $P$ and hence to
the value of the charm penguin $A_{c.c}$. In fact, for negligible
charm penguin, \ie, $A_{c.c}=0$ one finds that BRs are given by:%
\bea %
BR_{K^0\pi^+} &=& 2.1\times 10^{-6}, ~~~~~~~~~~~~~~~~ BR_{K^+\pi^-} =  2.3\times 10^{-6},\nonumber\\
BR_{K^+\pi^0} &=&  1.4\times 10^{-6}, ~~~~~~~~~~~~~~ BR_{K^0\pi^0} = 0.9\times 10^{-6}.%
\eea %
These results are smaller than the experimental measurements.
Therefore, it is appealing that large charm penguin is
essential for the consistency of the SCET. For the value of $A_{c.c.}$
in Eq.~(\ref{Acc}), one finds significant enhancement for the BRs
and they become close to the experimental results, namely they are
now given by:
\bea %
BR_{K^0\pi^+} &=& 20.5 \times 10^{-6}, ~~~~~~~~~~~~~~~~ BR_{K^+\pi^-} =  21.1\times 10^{-6},\nonumber\\
BR_{K^+\pi^0} &=&  11.2 \times 10^{-6}, ~~~~~~~~~~~~~~~ BR_{K^0\pi^0} = 9.7\times 10^{-6},%
\eea %

In order to understand the dependence of the CP asymmetries on
different contribution, we will neglect small $r_J^2$ corrections.
However our numerical results are based on the complete
expressions of the asymmetries, which turn out to be quite
close to the approximated ones. Keeping linear terms in $r_J$, one
finds that the $B\to K\pi$
CP asymmetries can be written as %
\begin{eqnarray} %
A^{CP}_{B^+\to K^0\pi^+} &=& \frac{ 2 r_A \sin{\delta_A}
\sin{\gamma}}{1 + 2 r_A \cos{\delta_A}
\cos{\gamma} },\nn\\
A^{CP}_{\bar{B^0}\to K^+\pi^-} &=& \frac{ 2 r_T \sin{\delta_T}
\sin{\gamma}}{1 +2 r^C_{EW} \cos{\delta^C_{EW}}+2 r_T
\cos{\delta_T}
\cos{\gamma}},\nn\\
 A^{CP}_{B^+\to K^+\pi^0} &=&\frac{2 r_T \sin{\delta_T} \sin{\gamma}+ 2 r_C \sin{\delta_C}
\sin{\gamma}+2 r_A \sin{\delta_A} \sin{\gamma}}{1 +2 r_{EW}
\cos{\delta_{EW}} +2 r_C \cos{\delta_C} \cos{\gamma}+2 r_T
\cos{\delta_T} \cos{\gamma}+2 r_A \cos{\delta_A}
\cos{\gamma}},\nn\\
A^{CP}_{B^0\to K^0\pi^0}&=&\frac{-2 r_C
\sin{\delta_C}\sin{\gamma}}{1
-2r_{EW}\cos{\delta_{EW}}+2r^C_{EW}\cos{\delta^C_{EW}}- 2 r_C
\cos{\delta_C}\cos{\gamma} }. %
\label{asymmetries1}
\end{eqnarray}

It is interesting to note that without charm penguin contribution,
although $r_J$ is not suppressed, all the CP asymmetries of $B\to
K \pi$ decays are quite small, ${\cal O}(0.01)$, which is not
consistent with the experimental results reported above in
Table~\ref{asymtable}. This is due to the lack of large strong
phases. As mentioned, the charm penguin in SCET is the main source
of strong phases. Therefore these phases associated with $r_J$ are
essentially given by $\pm 1/P$. This can be checked in
Eq.(\ref{r-SM}), where one observes
the following relation:%
\be%
\sin \delta_T = \sin \delta_C = \sin \delta_A = - \sin \delta_{EW}
= -\sin \delta_{EW}^C = -\sin \delta_P.%
\label{strong-phases}
\ee%
It is now clear that the above expression of the CP asymmetries
cannot lead to $A^{CP}_{K^+\pi^-}$ and $A^{CP}_{K^+ \pi^0}$ with
different sign. In fact, one can approximate these two asymmetries
as follows: $A^{CP}_{K^+ \pi^-} \simeq 2 r_T \sin \delta \sin
\gamma$ and $A^{CP}_{K^+ \pi^0} \simeq 2 (r_T+ r_C) \sin \delta
\sin \gamma$, which lies between $A^{CP}_{K^+\pi^-}$ and $2 A_{K^+
\pi^-}$. One can check this conclusion numerically. For instance,
with a charm penguin fixed by $B \to \pi \pi$ \cite{Bauer:2005kd},
one
finds the following asymmetries: %
\bea %
A^{CP}_{B^+\to K^0\pi^+} &=& - 0.01, ~~~~~~~~~~~~~~ A^{CP}_{B^0\to
K^+\pi^-} = - 0.03,\nonumber\\
A^{CP}_{B^+\to K^+\pi^0} &=& -0.04, ~~~~~~~~~~~~~~ A^{CP}_{B^0\to
K^0\pi^0} = 0.02.%
\eea %

Note that the EW penguins violate the isospin symmetry, hence they
are natural candidates for explaining the discrepancy between
$A^{CP}_{K^+\pi^-}$ and $A^{CP}_{K^+ \pi^0}$. However, as we have
seen, within the SM, these two asymmetries are not sensitive to the
values of $r_{EW}$ and $r_{EW}^c$. This is due to the fact that the
EW penguins are real in the SM and hence they have no interference
with the QCD penguin $P$. As emphasized in
Ref.~\cite{Khalil:2009zf}, a possible solution for the $B\to K \pi$
puzzle is to have a new source of CP violation that generates CP
phases for the EW penguins. This possibility can be implemented in
supersymmetric models and has been checked within the framework of
QCDF in Ref.~\cite{khalil:2005qg,Khalil:2009zf}.

\section{ \bf New Physics effects and CP asymmetries of {$ B\to K \pi$ }in SCET}
In this section we analyze the type of general NP beyond the SM
that can account for the CP asymmetries of $B\to K \pi$ and
explain the discrepancy between $A^{CP}_{K^+ \pi^-}$ and
$A^{CP}_{K^+ \pi^0}$. As mentioned above and discussed in detail
in Ref.~\cite{khalil:2005qg}, this NP must contain a new source of CP
violation beyond the CKM phase. The impact of any NP beyond the SM
appears only in the Wilson coefficients at electroweak scale.
Therefore, the total Wilson coefficients can be written as %
\be%
C_i = C_i^{SM} + C_i^{NP}~, ~~~~~~~~~~~~  i= 1,..,10,7\gamma,8g.%
\ee%
where $C_i^{NP}$ are generally complex, \ie, they have CP
violating phase, unlike the $C_i^{SM}$. Also the NP is expected to
have relevant contributions to the penguins and not to the tree
processes, which are dominated by the SM effects. Therefore, one
can assume that the color-tree and color suppressed parameters
remain as in the SM, \ie, $T = T^{SM}$ and $C =
C^{SM}$, while the penguin parameters are given by:%
\bea %
P e^{i\theta_P} e^{i\delta_P}&=& \vert P^{SM} \vert~
e^{i\delta_{c.c}} + \vert P^{NP} \vert~ e^{i\phi_P} = \vert P^{SM}
\vert \left[e^{i\delta_{c.c}} + \kappa_P e^{i\phi_P}\right], \\
P_{EW} e^{i\theta_{EW}}&=& \vert P_{EW}^{SM} \vert  + \vert
P_{EW}^{NP} \vert~ e^{i\phi_{EW}} = \vert
P_{EW}^{SM} \vert \left[1+ \kappa_{EW} e^{i\phi_{EW}}\right], \\
P_{EW^C} e^{i\theta^C_{EW}}&=& \vert P_{EW^C}^{SM} \vert + \vert
P_{EW^C}^{NP} \vert~ e^{i\phi_{EW^C}}=\vert
P_{EW^C}^{SM} \vert \left[1+ \kappa^C_{EW} e^{i\phi^C_{EW}}\right].%
\eea %
Here we assume that the only source of strong phase is
$\delta_{c.c}$ in $P^{SM}$. As mentioned in the previous section,
a large charm penguin contribution is very crucial in the SCET in
order to get the branching ratio of $B\to K \pi$ decays consistent
with the experimental measurements. Furthermore, it is also needed
to allow for a large strong phase, which is crucial for generating
a large CP asymmetry.  In order to generalize the parametrization
of $B \to K \pi$ in Eq.~(\ref{SMamplitude1}), one should rewrite $P$
as $P = \vert P \vert e^{i\delta_P} e^{i\theta_P}$, where
$\delta_P$ and $\theta_P$ are the strong (CP conserving) and CP
violating phases
associated with $P$, which can be determined as follows%
\bea%
\delta_{P} &=& \tan^{-1} \left(\frac{\sin \delta_{c.c}}{\cos
\delta_{c.c} + \kappa_P \cos \phi_P}\right), ~~~~~~~~~~~\theta_{P}
= \tan^{-1} \left(\frac{\kappa_P \sin\phi_P}{\kappa_P
\cos \phi_P +  \cos \delta_{c.c}}\right).%
\eea%
Similarly, $\theta_{EW}$ and $\theta_{EW}^C$ can be defined in
terms of $\phi_{EW}$ and $\phi_{EW}^C$. In this case, the ratio between the EW and QCD penguins can be written as:%
\be %
\frac{P_{EW}}{P} = r_{EW} e^{-i \delta_{P}} e^{i
(\theta_{EW}-\theta_P)},
~~~~~~~~~~ \frac{P_{EW}^C}{P} = r_{EW}^C e^{-i \delta_{P}} e^{i (\theta_{EW}^C- \theta_P)}, %
\ee %
where $r_{EW}$ and $r_{EW}^C$ are given by%
\be%
r_{EW} = (r_{EW})^{SM} \left\vert \frac{1+ \kappa_{EW} e^{i
\phi_{EW}}}{1+ \kappa_{P} e^{i(\phi_P -
\delta_{c.c})}}\right\vert, ~~~~~~~~~~~r_{EW}^C = (r_{EW}^C)^{SM}
\left\vert \frac{1+ \kappa_{EW}^C e^{i \phi_{EW}^C}}{1+ \kappa_{P}
e^{i(\phi_P -
\delta_{c.c})}}\right\vert.%
\ee %
Note that the strong phases still satisfy the relation in
Eq.~(\ref{strong-phases}), as in the SM. This leads to the
following parametrization for the $B \to K \pi$ amplitudes:
\begin{eqnarray}%
\label{Kpigeneral}%
A(B^+ \to K^0  \pi^+ )& =& \lambda_c^{(s)} P \Big[ e^{i \theta_P}
+ r_A e^{i(\delta_A -\gamma)}  \Big]\,, \nn \\
 A(B^0 \to K^+ \pi^-) & =&  \lambda_c^{(s)} P \Big[ e^{i \theta_P} +
\big( r^C_{EW} e^{i(\theta_{EW}^C+\delta^C_{EW})} + r_T e^{i(\delta_T-\gamma)}\big) \Big], \nn \\
 \sqrt{2} A(B^+ \to K^+ \pi^0)&=&  \lambda_c^{(s)} P \Big[ e^{i \theta_P} +
 \big( r_{EW} e^{i(\theta_{EW}+\delta_{EW})} + r_T e^{i(\delta_T-\gamma)}
+ r_C e^{i(\delta_C-\gamma)}+ r_A e^{i(\delta_A-\gamma)}  \big)  \Big], \nn \\
\sqrt{2} A(B^0 \to K^0 \pi^0) &=& \lambda_c^{(s)} P \Big[ - e^{i
\theta_P} + \big( r_{EW} e^{i(\theta_{EW}+\delta_{EW})} - r^C_{EW}
e^{i(\theta_{EW}^C +\delta_{EW}^C)} + r_C e^{i(\delta_C-\gamma)}
\big)
  \Big].
\label{SMamplitude}
\end{eqnarray}%
In this case, one finds that the approximate expressions for the
CP asymmetries in Eq.(\ref{asymmetries1}) can be generalized as
follows: %
\begin{eqnarray} %
A^{CP}_{B^+\to K^0\pi^+} &=& \frac{ 2 r_A \sin{\delta_A} \sin(
\theta_P+ \gamma)}{1 + 2 r_A \cos{\delta_A}
\cos(\theta_P+\gamma) },\nn\\
A^{CP}_{\bar{B^0}\to K^+\pi^-} &=& \frac{ 2 r_T \sin{\delta_T}
\sin(\theta_P + \gamma) + 2 r_{EW}^C \sin{\delta_{EW}^C}
\sin(\theta_P - \theta_{EW}^C)}{1 +2 r_T \cos{\delta_T}
\cos(\theta_P+\gamma)+2 r^C_{EW} \cos{\delta^C_{EW}} \cos(\theta_P - \theta_{EW}^C)},\nn\\
A^{CP}_{B^+\to K^+\pi^0} &=&\frac{2 r_{EW}
\sin{\delta_{EW}}\sin(\theta_P -\theta_{EW})+ 2 \left[r_T
\sin{\delta_T} +r_C \sin{\delta_C} + r_A \sin{\delta_A}\right]
\sin(\theta_P +\gamma)}{1 +2 r_{EW} \cos{\delta_{EW}}\cos(\theta_P
-\theta_{EW})+ 2 \left[r_T \cos{\delta_T} +r_C \cos{\delta_C} +
r_A \cos{\delta_A}\right]
\cos(\theta_P +\gamma)},\nn\\
A^{CP}_{B^0\to K^0\pi^0}&=&\frac{-2 r_{EW} \sin \delta_{EW}
\sin(\theta_P - \theta_{EW}) + 2 r_{EW}^C \sin \delta_{EW}^C
\sin(\theta_P - \theta_{EW}^C)-2 r_C
\sin{\delta_C}\sin(\theta_P+\gamma)}{1 -2r_{EW}\cos{\delta_{EW}}
\cos(\theta_P -\theta_{EW})+2r^C_{EW}\cos{\delta^C_{EW}}
\cos(\theta_P -\theta_{EW}^C)- 2 r_C
\cos{\delta_C}\cos{\theta_P +\gamma}}.\nn\\ %
\label{asymmetries2}
\end{eqnarray}

If one assumes $r_C \sim r_T$, and neglect the small $r_A$, then
the CP asymmetries $A^{CP}_{K^+\pi^-}$ and $A^{CP}_{K^+ \pi^0}$,
which are not consistent with the SM
results, can be written as:%
\bea %
A^{CP}_{K^+ \pi^-} &\simeq& \frac{ 2\sin \delta_P\left[- r_T
\sin(\theta_P + \gamma) + r_{EW}^C  \sin(\theta_P
-\theta_{EW}^C)\right]}{1+ 2 r_T \cos \delta_P \cos(\theta_P +
\gamma)+
2 r_{EW}^C \cos \delta_P \cos(\theta_P - \theta_{EW}^C)}, \nn\\
A^{CP}_{K^+ \pi^0} &\simeq & \frac{ 2 \sin \delta_P\left[ r_{EW}
\sin(\theta_P -\theta_{EW}) - 2 r_T \sin(\theta_P+\gamma)\right]
}{1+ 2 r_{EW} \cos \delta_P \cos(\theta_P
-\theta_{EW}) + 4 r_T \cos\delta_P \cos(\theta_P+\gamma)} . %
\label{asym2}%
\eea %
Therefore, the difference between these two asymmetries is now given by%
\be %
A^{CP}_{K^+ \pi^0} - A^{CP}_{K^+ \pi^-}\simeq 2 \sin \delta_P
\left[ r_{EW}\sin(\theta_P - \theta_{EW}) -  r_T \sin(\theta_P +
\gamma) - r_{EW}^C \sin(\theta_P - \theta_{EW}^C)\right] . %
\ee%
Note that the denominators in Eq.(\ref{asym2}) can be approximated
to one if large phases are considered to maximize the asymmetries.
According to Eq.(\ref{difference}), this difference should be of
order ${\cal O}(0.14)$ in order to match
the current experimental results. Thus one finds %
\be %
r_{EW} \sin(\theta_P -\theta_{EW}) - r_{T} \sin(\theta_P+\gamma)
-r_{EW}^C \sin(\theta_P - \theta_{EW}^C) \simeq \frac{0.07}{\sin \delta_P}. %
\ee %
Moreover, the result of $A^{CP}_{K^+
\pi^-}$ implies that  %
\be %
- r_T \sin (\theta_P + \gamma) + r_{EW}^C \sin(\theta_P
-\theta_{EW}^C) \sim \frac{- 0.049}{\sin\delta_P}.%
\ee %
{}From these relations, one gets:%
\be %
r_{EW} \sin(\theta_P -\theta_{EW}) - 2 r_{EW}^C \sin(\theta_P
-\theta_{EW}^C) \simeq \frac{0.12}{\sin\delta_P}. \label{condEW}
\ee%
This condition can be fulfilled if one of the following scenarios
takes place:\\
\begin{itemize}
\item $r_{EW} \sin(\theta_P - \theta_{EW}) \sim 0.12/\sin
\delta_P$, while $r_{EW}^C \sin(\theta_P - \theta_{EW}^C)\lsim
{\cal O}(0.01)$, which could be due to smallness of $r_{EW}^C$ or
$\theta_P \sim \theta_{EW}^C$. Note that if $\delta_{P} \sim
\delta_{c.c}$, then $r_{EW} \sin(\theta_P -\theta_{EW}) \sim 0.3$.
In this case, the required NP should enhance the value of $r_{EW}$
to be larger than $\vert 0.12/\sin\delta_P \vert$ and induce CP
violating phases such that $\sin(\theta_P -\theta_{EW}) \sim {\cal
O}(1)$, \ie, $\theta_{EW} \simeq \theta_{P} - \pi/2$. The phase
$\theta_P$ can be fixed from $A^{CP}_{K^+\pi^-}$ which in this
scenario is given by $2 r_T \sin \delta_P \sin(\theta_P + \gamma)$.

\item $r_{EW}\sim r_{EW}^C$ and $ \theta_{EW} \sim\theta_{EW}^C$ .
In this case, the required NP should lead to $r_{EW}\sin(\theta_P
-\theta_{EW}) \sim r_{EW}^C\sin(\theta_P -\theta_{EW}^C) \sim -
0.12/\sin\delta_P$. Therefore, $r_{EW}$ should be also larger that
$\vert 0.12/\sin \delta_P \vert$ and $\sin(\theta_P
-\theta_{EW}^C) \sim {\cal O}(-1)$, \ie, $\theta_{EW} \sim
\theta_{EW}^C \sim \theta_P +\pi/2$.

\item Another possibility is that $r_{EW} \sin (\theta_P
-\theta_{EW}) \lsim {\cal O}(0.01)$ and $r_{EW}^c \sin (\theta_P
-\theta_{EW^C}) \sim - 0.06/\sin\delta_P$. It may be natural to
think that color allowed contribution should dominate the color
suppressed ones, therefore this scenario requires a NP that
implies: $\theta_P \sim \theta_{EW}$ and $\sin (\theta_P
-\theta_{EW}^C) \sim - 0.06/(r_{EW}^C \sin\delta_P)$

\end{itemize}

It is important to note that in these three marked scenarios, the
new CP violating phases are crucial and play important role in
modifying the $B \to K \pi$ CP asymmetries and moving them towards
the experimental measurements. This could be an interesting test
for the correct NP that we should consider as extension of the SM.
In the next section we will check the possibility that SUSY can
resolve the puzzle of $B \to K \pi$ as it can do in the QCDF
\cite{khalil:2005qg} and if it is so, which scenario of the above
three can be implemented in SUSY models. It is also worth
mentioning that if the denominators of Eq. (\ref{asymmetries2})
are less than one, then the value of the CP asymmetries can be
enhanced and smaller values of CP phases could be sufficient for
accommodating the experimental results of CP asymmetries of $B\to
K \pi$ decays.

Before concluding this section, it is worth mentioning that in
QCDF there are more than one source of strong phases, therefore
one may adjust the sign of $\delta_{EW}$ and $\delta^C_{EW}$ such
that the difference between $A^{CP}_{K^+ \pi^-}$ and $A^{CP}_{K^+
\pi^0}$ can be obtained without any tight relation between the CP
violating phases of the QCD and EW penguins, like those obtained
in SCET. Accordingly, it is expected to be more difficult for NP
to account for the CP asymmetry of $B\to K \pi$ decays in SCET
than in other frames of hadron dynamics.

\section{ \bf SUSY contributions to the CP asymmetry of
{$ B\to K \pi$ }in SCET}

Now, we consider SUSY as a potential candidate for NP beyond the
SM and analyze its contribution to the CP asymmetry of $ B\to K
\pi$ in SCET. As mentioned, the impact of SUSY appears only in the
Wilson coefficients at the electroweak scale. Here we focus on the
relevant contributions that may play important role in the CP
asymmetry of $B \to K \pi$, in particular the gluino contribution
to the chromomagnetic and EW penguins, namely
$C_{8g}^{\tilde{g}}$, $C_7^{\tilde{g}}$ and $C_9^{\tilde{g}}$, and
in addition, the chargino contribution to the $Z$-penguin
$C_9^{\chi}$.
These can be written in MIA as \cite{Gabrielli:2004yi,Lunghi:1999uk}:%
\bea %
C_{8g}^{\tilde{g}} &\simeq& \frac{8\alpha_S \pi}{9 \sqrt{2} G_F
V_{tb} V^*_{ts} m_{\tilde{q}}^2} \frac{m_{\tilde{g}}}{m_b}
\left[\left(\delta^d_{LR}\right)_{23}
+\left(\delta^d_{RL}\right)_{23}\right]\left(\frac{1}{3} M_1(x)+ 3
M_2(x)\right),  \\
C_{7\gamma}^{\tilde{g}} &\simeq& \frac{\pi \alpha_S}{6\sqrt{2} G_F
V_{tb} V^*_{ts} m_{\tilde{q}}^2}\frac{N_c^2 -1}{2 N_c} \left[
\left(\delta^d_{LL}\right)_{23} \frac{1}{4} P_{1,3,2}(x,x) +
\left(\delta^d_{RL}\right)_{23} \frac{m_{\tilde{g}}}{m_b}
P_{1,2,2}(x,x) \right],\\
C_{9}^{\tilde{g}} &\simeq& -\frac{\pi \alpha_S}{6\sqrt{2} G_F
V_{tb} V^*_{ts} m_{\tilde{q}}^2}\frac{N_c^2 -1}{2 N_c}
\left(\delta^d_{LL}\right)_{23} \frac{1}{3} P_{0,4,2}(x,x) ,\\
C_{9}^{\chi} &\simeq& \frac{\alpha}{4 \pi} Y_t \left[
\left(\delta^u_{RL}\right)_{32}+\lambda
\left(\delta^u_{RL}\right)_{31}\right] \left(4 (1-\frac{1}{4\sin^2\theta_W}) R_C + R_D \right). %
\eea%
where $x=m_{\tilde{g}}^2/m_{\tilde{q}}^2$ and the functions
$M_{1,2}$, $P_{ijk}$ and $R_{C,D}$ are the corresponding loop
functions, which depend on SUSY parameters through gluino/chargino
mass and squark masses and can be found in
Ref.~\cite{Gabrielli:2004yi,Lunghi:1999uk}. Note that although
$(\delta^d_{LR})_{23}$ and $(\delta^d_{RL})_{23}$ are constrained
by the experimental limits of $b \to s \gamma$ to be less than
${\cal O}(10^{-2})$, their contributions to $C_{8g}^{\tilde{g}}$
and $C_{7}^{\tilde{g}}$ are enhanced by a large factor of
$m_{\tilde{g}}/m_b$. On the other hand, the mass insertion
$(\delta^u_{RL})_{32}$ is free from any stringent constraints, and
it can be of order one.

As advocated in the introduction, SUSY models include new CP
violating phases beyond the SM phase $\delta_{CKM}$. These phases
arise from the complex soft SUSY breaking terms. In MIA, the SUSY
CP violating phases lead to complex mass insertions
$(\delta^{u,d}_{AB})_{ij}$, hence complex SUSY Wilson
coefficients, unlike in SM. A SUSY model with non-universal
$A$-terms, which can be obtained in most of SUSY breaking
scenarios is the natural framework for inducing new SUSY sources
of CP and flavor violation that yield observable effects in the
low –energy CP violation experiments without exceeding the
experimental EDM limits \cite{Abel:2001vy}. For $m_{\tilde{g}}
=300$ GeV and $m_{\tilde{q}}=500$ GeV, the SUSY contributions to QCD and EW penguins can be approximated by%
\bea %
(\hat{P})_{SUSY}&=& (-0.004 + 0.0002 i) (\delta^ d_{LL})_{23} -
0.36 (\delta^d_{LR})_{23} -  0.36 (\delta^d_{RL})_{23}
- 0.00004 (\delta^u_{RL})_{32},\\
(\hat{P}_{EW})_{SUSY}&=& (0.025 - 0.0005 i) (\delta^ d_{LL})_{23} + 0.00031 (\delta^u_{RL})_{32},\\
(\hat{P}^C_{EW})_{SUSY}&=& (0.013 - 0.0005 i) (\delta^ d_{LL})_{23} + 0.00013 (\delta^u_{LR})_{32}. %
\eea %
Recall that the SM contribution to these parameters are given by%
\bea%
(\hat{P})_{SM}&=& -0.006 + 0.0016 i ,\\
(\hat{P}_{EW})_{SM}&=& -0.0005 + 0.0001i,\\
(\hat{P}^C_{EW})_{SM}&=& -0.0002 + 0.0001 i. %
\eea%

>From the $b\to s \gamma$ constraints, one can fix the relevant
mass insertions as follows: %
\be%
(\delta^ d_{LL})_{23} = e^{i \alpha^d_1},~~~~~~~~~ (\delta^
d_{LR})_{23}=(\delta^ d_{RL})_{23} = 0.01
e^{i\alpha^d_2},~~~~~~~~~
(\delta^ u_{RL})_{32} = 1 e^{i \alpha^u}~,%
\ee%
with unconstrained CP violating phases: $\alpha^d_{1,2}$ and
$\alpha^u$. It is clear that the QCD penguin is dominated by the SM
contribution, which is essentially the charm penguin effect.
However, the EW penguins, which are quite suppressed in the SM,
receive significant contributions in the SUSY models, in particular due
to the gluino contribution to EW penguin with photon mediation.
In this case, one can approximate $r_{EW}$ and
$r_{EW}^C$ as%
\bea%
r_{EW} &=& (r_{EW})^{SM} \left\vert \frac{1- 46.5 e^{i \alpha^d_1}
- 0.58 e^{i \alpha^u}}{1+ (0.65+0.13~i) e^{i \alpha^d_1}+
(1.64+0.41~i) e^{i \alpha^d_2}}\right\vert ,\l{rEW}\\
r_{EW}^C &\simeq& (r_{EW}^C)^{SM} \left\vert \frac{1- 50.7 e^{i
\alpha^d_1} - 0.52 e^{i \alpha^u}}{1+ (0.65+0.13~i) e^{i
\alpha^d_1}+ (1.64+0.41~i) e^{i \alpha^d_2}}\right\vert  . %
\eea %
{}From these expressions, it is clear that the magnitudes of
$r_{EW}$ and $r_{EW}^C$ can be significantly enhanced and reach up
to tens of the SM results. As we concluded in the previous
section, a large value of $r_{EW}$ and/or $r_{EW}^C$, besides
non-vanishing CP violating phases $\theta_{EW}$ and
$\theta_{EW}^C$, is an essential condition for resolving the $B
\to K \pi$ puzzle. Also one notes that the chargino exchange gives
sub-dominant contribution.

One can also notice that the relation $r_{EW} \sim 2 r_{EW}^C$
remains valid in SUSY models, as in the SM. Furthermore, since the
mass insertion $(\delta_{LL}^d)_{23}$ gives the dominant
contributions to $P_{EW}$ and $P_{EW}^C$, one gets $\sin(\theta_P
-\theta_{EW}) \sim \sin(\theta_P -\theta_{EW}^C)$. Therefore, the
condition of accounting for the discrepancy in $B\to K \pi$ CP
asymmetries, Eq.(\ref{condEW}), leads to %
\be%
r_{EW} \sin(\theta_P -\theta_{EW}) \left(1 -\frac{2
r_{EW}^C}{r_{EW}}\right) \simeq \frac{0.12}{\sin \delta_P} \sim
0.4,%
\ee %
where $\sin(\theta_P -\theta_{EW}) \sim {\cal O}(1)$ and $(1 -2
r_{EW}/r_{EW}^C) \sim {\cal O}(0.1)$. Therefore, the CP
asymmetries of $B \to K \pi$ can be accommodated if $r_{EW} \geq
{\cal O}(1)$, which can be obtained as shown in Eq.(\ref{rEW}).

As an example, one can check that the following values of the mass
insertion phases: $\alpha_1^d = 2.1$ rad, $\alpha_2^d = 1.5$ rad,
and $\alpha^u =0$ lead to $r_{EW} \simeq 1.7$ and $r_{EW}^C=0.9$.
This means that both $r_{EW}$ $r^C_{EW}$ are enhanced from the SM
result by a factor of twenty. Also, in this case, one finds the
SUSY CP violating phases as follows: $\theta_{EW}=-2.25$ rad and
$\theta_{EW}^C= -2.27$ rad. These results imply that the CP
asymmetries of $B \to K^+
\pi^0$ and $B \to K^+ \pi^-$ are given by%
\be%
A^{CP}_{K^+ \pi^0} = 0.06, ~~~~~~~~~~~~ A^{CP}_{K^+ \pi^-} =
-0.09,%
\ee%
which are in agreement with the experimental measurements reported
in Table 1. It is important to note that since $r_{EW} << 1$, one
must use the complete expression for the CP asymmetries to get the
correct results.

\section{ \bf Conclusions}

In this work, we have studied the large discrepancy in the
experimentally measured asymmetries of $B\rightarrow K\pi$ in the
SCET framework. We conclude that in the Standard Model, one
cannot accommodate all the experimental results in the SCET framework.

We have considered the possibility that New Physics could
satisfy the measured asymmetries.
We have classified the properties of New Physics needed to bring the
theoretical results to experimentally acceptable level in the SCET
scenario.  A general feature is that a new source of CP violation
must emerge.
As an example of a New Physics model, we studied supersymmetric
models with minimal particle content in
a model independent fashion by utilizing mass insertion approximation.
We found that the gluino contribution to the electroweak penguin is
essential.
In our analysis we let trilinear $A$-terms vary freely, in which case
we can find an experimentally allowed region in the parameter space.

Therefore, if SCET is a reliable way to treat hadronic matrix elements,
the present experimental results indicate New Physics.
Supersymmetric models remain a viable candidate for such
New Physics, if nonminimal flavor violation is allowed.

\section*{ \bf Acknowledgements}

We would like to thank G. Faisel for collaboration. K.H. is
grateful for the support by the Academy of Finland (Project No.
115032). S.K. is grateful for the support by the Science and
Technology Development Fund (STDF) Project ID 437, the academy of
scientific research and technology, and the ICTP Project ID 30.



\begin{thebibliography}{99}

\bibitem{Bona:2008jn}
  M.~Bona {\it et al.}  [UTfit Collaboration],
Transitions,''
  arXiv:0803.0659 [hep-ph].
\bibitem{HFAG}
http://www.slac.stanford.edu/xorg/hfag/, [Heavy Flavor Averaging
Group (HFAG) Collaboration].
\bibitem{beneke}
M.~Beneke, G.~Buchalla, M.~Neubert and C.T.~Sachrajda, Phys.\
Rev.\ Lett.\  {\bf 83}, 1914 (1999)[arXiv:hep-ph/9905312]; Nucl.\
Phys.\  B {\bf 606}, 245 (2001)[arXiv:hep-ph/0104110]
\bibitem{khalil:2005qg}
  S.~Khalil,
  Phys.\ Rev.\  D {\bf 72}, 035007 (2005)
  [arXiv:hep-ph/0505151]
\bibitem{Bauer:2000ew}
  C.~W.~Bauer, S.~Fleming and M.~E.~Luke,
  Phys.\ Rev.\  D {\bf 63}, 014006 (2000)
  [arXiv:hep-ph/0005275].
\bibitem{Bauer:2005kd}
  C.~W.~Bauer, I.~Z.~Rothstein and I.~W.~Stewart,
  Phys.\ Rev.\  D {\bf 74}, 034010 (2006)
\bibitem{Bauer:2000yr}
  C.~W.~Bauer, S.~Fleming, D.~Pirjol and I.~W.~Stewart,
  Phys.\ Rev.\  D {\bf 63}, 114020 (2001)
  [arXiv:hep-ph/0011336].
  \bibitem{Chay:2003zp}
  J.~g.~Chay and C.~Kim,
  Phys.\ Rev.\ D {\bf 68}, 071502 (2003)
  [arXiv:hep-ph/0301055].
\bibitem{Chay:2003ju}
  J.~Chay and C.~Kim,
  Nucl.\ Phys.\ B {\bf 680}, 302 (2004)  [arXiv:hep-ph/0301262].\\
  D.~Pirjol and I.~W.~Stewart,
  Phys.\ Rev.\  D {\bf 67}, 094005 (2003)
  [Erratum-ibid.\  D {\bf 69}, 019903 (2004)]
  [arXiv:hep-ph/0211251].\\
\bibitem{Jain:2007dy}
  A.~Jain, I.~Z.~Rothstein and I.~W.~Stewart,
  arXiv:0706.3399 [hep-ph].\\
  M.~Beneke and S.~Jager,
  Nucl.\ Phys.\  B {\bf 751}, 160 (2006)
  [arXiv:hep-ph/0512351].\\
  M.~Beneke and S.~Jager,
  Nucl.\ Phys.\  B {\bf 768}, 51 (2007)
  [arXiv:hep-ph/0610322].\\
  C.~W.~Bauer, D.~Pirjol and I.~W.~Stewart,
  Phys.\ Rev.\ D {\bf 65}, 054022 (2002)
  [arXiv:hep-ph/0109045].\\
  C.~W.~Bauer, S.~Fleming, D.~Pirjol, I.~Z.~Rothstein and I.~W.~Stewart,
  Phys.\ Rev.\ D {\bf 66}, 014017 (2002).
  [arXiv:hep-ph/0202088].
 \bibitem{Bauer:2002aj}
  C.~W.~Bauer, D.~Pirjol and I.~W.~Stewart,
  Phys.\ Rev.\  D {\bf 67}, 071502 (2003)
  [arXiv:hep-ph/0211069].
\bibitem{Khalil:2009zf}
  S.~Khalil, A.~Masiero and H.~Murayama,
  arXiv:0908.3216 [hep-ph], to appear in Phys. Lett. B.
\bibitem{Abel:2001vy}
  S.~Abel, S.~Khalil and O.~Lebedev,
  Nucl.\ Phys.\  B {\bf 606}, 151 (2001)
  [arXiv:hep-ph/0103320].
\bibitem{Buchalla:1995vs}
  G.~Buchalla, A.~J.~Buras and M.~E.~Lautenbacher,
  Rev.\ Mod.\ Phys.\  {\bf 68}, 1125 (1996)
  [arXiv:hep-ph/9512380].
\bibitem{Bauer:2004tj}
  C.~W.~Bauer, D.~Pirjol, I.~Z.~Rothstein and I.~W.~Stewart,
  Phys.\ Rev.\ D {\bf 70}, 054015 (2004)
  [arXiv:hep-ph/0401188].\\
  M.~Beneke, G.~Buchalla, M.~Neubert and C.~T.~Sachrajda,
  Nucl.\ Phys.\  B {\bf 591}, 313 (2000)
  [arXiv:hep-ph/0006124].
\bibitem{Keum:2000ph}
  Y.~Y.~Keum, H.~n.~Li and A.~I.~Sanda,
  Phys.\ Lett.\  B {\bf 504}, 6 (2001)
  [arXiv:hep-ph/0004004].
\bibitem{Lu:2000em}
  C.~D.~Lu, K.~Ukai and M.~Z.~Yang,
  Phys.\ Rev.\  D {\bf 63}, 074009 (2001)
  [arXiv:hep-ph/0004213].
\bibitem{Beneke:2001ev}
  M.~Beneke, G.~Buchalla, M.~Neubert and C.~T.~Sachrajda,
  Nucl.\ Phys.\  B {\bf 606}, 245 (2001)
  [arXiv:hep-ph/0104110].
\bibitem{Kagan:2004uw}
  A.~L.~Kagan,
  Phys.\ Lett.\  B {\bf 601}, 151 (2004)
  [arXiv:hep-ph/0405134].
\bibitem{Arnesen:2006vb}
  C.~M.~Arnesen, Z.~Ligeti, I.~Z.~Rothstein and I.~W.~Stewart,
  arXiv:hep-ph/0607001.
\bibitem{Williamson:2006hb}
  A.~R.~Williamson and J.~Zupan,
  Phys.\ Rev.\  D {\bf 74}, 014003 (2006)
  [Erratum-ibid.\  D {\bf 74}, 03901 (2006)]
  [arXiv:hep-ph/0601214].
\bibitem{Gabrielli:2004yi}
  E.~Gabrielli, K.~Huitu and S.~Khalil,
  Nucl.\ Phys.\  B {\bf 710}, 139 (2005)
  [arXiv:hep-ph/0407291].
\bibitem{Lunghi:1999uk}
  E.~Lunghi, A.~Masiero, I.~Scimemi and L.~Silvestrini,
  Nucl.\ Phys.\  B {\bf 568}, 120 (2000)
  [arXiv:hep-ph/9906286].






\end{thebibliography}
\end{document}